\documentclass{article}

\PassOptionsToPackage{numbers, compress}{natbib}
\usepackage[preprint]{neurips_2024}

\usepackage[utf8]{inputenc} 
\usepackage[T1]{fontenc}    
\usepackage[colorlinks,
            linkcolor=red,
            anchorcolor=blue,
            citecolor=green
            ]{hyperref}
\usepackage{url}            
\usepackage{booktabs}       
\usepackage{amsfonts}       
\usepackage{nicefrac}       
\usepackage{microtype}      
\usepackage{xcolor}         
\usepackage{caption}
\usepackage{subcaption}
\usepackage{natbib}
\usepackage{graphicx}
\usepackage{enumitem}
\usepackage{makecell}
\usepackage{amsmath}
\usepackage{amsthm}
\usepackage{thmtools}
\usepackage{thm-restate}
\usepackage{algorithm}
\usepackage{algorithmic}
\usepackage{bbm}
\usepackage{tabularx}
\usepackage{listings}
\usepackage{xcolor}
\usepackage{multirow}
\usepackage{multicol}
\usepackage{tcolorbox}
\usepackage{tikz}
\usepackage{footnote}
\usepackage{wrapfig}

\definecolor{tblue}{RGB}{31,119,180}
\definecolor{torange}{RGB}{255,127,14}
\definecolor{tgreen}{RGB}{44,160,44}
\definecolor{tred}{RGB}{214,39,40}
\definecolor{tpurple}{RGB}{148,103,189}
\definecolor{lightblue}{RGB}{173, 216, 230}
\definecolor{lightpink}{RGB}{255, 182, 193}
\definecolor{lightgreen}{RGB}{144, 238, 144}

\usepackage{colortbl}
\usepackage{xcolor}
\usepackage{array}

\newcommand{\hide}[1]{} 

\newcommand{\eg}{\textit{e}.\textit{g}.}

\def\model{VideoRAG}

\title{VideoRAG: Retrieval-Augmented Generation with Extreme Long-Context Videos}

\author{
  Xubin Ren\textsuperscript{1}\thanks{Equal contribution.}\footnotemark[1] ~~~
  Lingrui Xu\textsuperscript{1}\footnotemark[1] ~~~
  Long Xia\textsuperscript{2} ~~~
  Shuaiqiang Wang\textsuperscript{2} ~~~
  Dawei Yin\textsuperscript{2} ~~~
  Chao Huang\textsuperscript{1}\thanks{Chao Huang is the Corresponding Author.} \\
  \textsuperscript{1}The University of Hong Kong ~~~
  \textsuperscript{2}Baidu Inc. \\
  \texttt{\{xubinrencs, lingruixu.db, long.phil.xia, chaohuang75\}@gmail.com} \\
  \texttt{wangshuaiqiang@baidu.com yindawei@acm.org}
}

\begin{document}

\maketitle

\begin{abstract}
Retrieval-Augmented Generation (RAG) has demonstrated remarkable success in enhancing Large Language Models (LLMs) through external knowledge integration, yet its application has primarily focused on textual content, leaving the rich domain of multi-modal video knowledge predominantly unexplored. This paper introduces \model, the first retrieval-augmented generation framework specifically designed for processing and understanding extremely long-context videos. Our core innovation lies in its dual-channel architecture that seamlessly integrates (i) graph-based textual knowledge grounding for capturing cross-video semantic relationships, and (ii) multi-modal context encoding for efficiently preserving visual features. This novel design empowers \model\ to process unlimited-length videos by constructing precise knowledge graphs that span multiple videos while maintaining semantic dependencies through specialized multi-modal retrieval paradigms. Through comprehensive empirical evaluation on our proposed LongerVideos benchmark-comprising over 160 videos totaling 134+ hours across lecture, documentary, and entertainment categories-\model\ demonstrates substantial performance compared to existing RAG alternatives and long video understanding methods. The source code of \model\ implementation and the benchmark dataset are openly available at: \textcolor{blue}{\url{https://github.com/HKUDS/VideoRAG}}.
\end{abstract}


\section{Introduction}\label{sec:intro}

Recent advances in Large Language Models (LLMs) have revolutionized NLP, yet their performance is inherently limited by the knowledge captured during pre-training. To address this limitation, Retrieval-Augmented Generation (RAG) has emerged as a powerful paradigm that enhances LLMs by dynamically retrieving and incorporating external knowledge during inference~\cite{ChunkRAG,gutierrez2024hipporag}. While RAG has demonstrated success across various text-based applications, such as question answering, and factual reasoning, its potential remains largely untapped in the rich domain of multi-modal content, particularly video understanding. The extension of RAG to video content presents unique challenges and opportunities, as videos contain complex multi-modal features, temporal dynamics, and intricate semantic relationships that go beyond traditional text-based knowledge integration approaches.

Although large vision models have achieved impressive progress in video understanding tasks, they face limitations when processing long-context videos. These models (\eg, VideoLLaMA3~\cite{VideoLLaMA3} and LLaVA-Video~\cite{LLaVA-Video}), primarily designed for short video clips, struggle to effectively capture and reason about temporal dependencies spanning multiple hours. The challenge becomes particularly acute in scenarios requiring cross-video understanding and knowledge integration, such as lecture series comprehension, documentary analysis, or sequential entertainment content interpretation. Current approaches often fragment long videos into isolated clips, leading to loss of contextual information and inability to establish meaningful connections across different videos. This limitation severely hampers applications in educational content analysis, media archiving, and video-based knowledge extraction, where understanding the broader context across multiple videos is essential.

The key challenges in realizing Retrieval-Augmented Generation for extreme long-context videos are multifaceted. (i) \textbf{Capturing Heterogeneous Video Knowledge}. Videos contain rich information across multiple modalities, including visual frames, audio streams, and textual descriptions. Effectively capturing and organizing this diverse knowledge presents a unique challenge that cannot be addressed by existing text-based RAG approaches. Existing methods are ill-equipped to handle the complexity of cross-modal information and their relationships. (ii) \textbf{Preserving Semantic Coherence for Cross-Video Understanding}. Maintaining the semantic connections across numerous videos, which may span hours or days, is more complex than a single video. Preserving these intricate relationships and comprehensive knowledge interdependencies is crucial for holistic video understanding. (iii) \textbf{Efficient Video Knowledge Retrieval}. When the video knowledge base consists of an unrestricted number of lengthy videos, quickly identifying the most pertinent clips in response to user queries becomes significantly more challenging. The retrieval system must provide users with the most relevant information to answer queries accurately.

By addressing these key challenges, the \model\ aims to unlock the full potential of RAG in the domain of extreme long-context videos, enabling powerful and comprehensive video understanding capabilities. At the heart of VideoRAG are two interlocking components - the \textbf{Multi-Modal Video Knowledge Indexing} framework and the \textbf{Knowledge-Grounded Multi-Modal Retrieval} paradigm. The indexing framework transforms video content into structured textual and visual representations, with graph-based textual knowledge grounding to preserve semantic relationships across videos, complemented by multi-modal context encoding to capture fine-grained cross-modal interactions. This dual-channel architecture enables VideoRAG to effectively organize and index long-context videos, preserving the rich semantics of the multimedia content. The knowledge-grounded retrieval paradigm then integrates textual semantic and visual content matching, leveraging the indexed knowledge graph and embeddings to identify the most relevant video content. Finally, VideoRAG employs a two-stage content extraction process that combines LLM-powered keyword extraction and vision-language model-based text grounding to enrich the visual analysis with text-based retrieval, generating comprehensive outputs for the final response.

The comprehensive evaluation on the benchmark datasets demonstrates the advantages and effectiveness of the \model\ framework in understanding extremely long-context videos, going beyond the limitations of existing RAG alternatives and large vision models. The results showcase \model's superior performance in effectively organizing and indexing long-form video content, allowing for precise retrieval of relevant segments across different video sources in response to user queries. Our ablation studies provide deeper insights into the individual contributions of \model's key components, highlighting the importance of the graph-based knowledge grounding and multi-modal retrieval mechanisms in elevating its performance. Furthermore, case studies demonstrate \model's practical applications in real-world scenarios, such as video-based knowledge extraction and educational content analysis, unlocking new possibilities for cross-video comprehension.

Moreover, the proposed LongerVideos curates a diverse collection of over 160 long-form videos spanning 134+ hours across lecture, documentary, and entertainment categories - a substantial advancement over existing datasets that are limited to inference on single~\cite{MLVU, LVBench} or relatively short-video content~\cite{Video-MME, EgoSchema}. LongerVideos enables the assessment of models' capabilities in reasoning across multiple long-context videos, a crucial requirement for real-world cross-video understanding scenarios, like video-based knowledge extraction and educational content analysis. By providing a robust testbed for evaluating long video understanding methods, this benchmark will advance the development of systems that can comprehend and reason about long-form video content at scale.
\section{Preliminary}\label{sec:preliminary}

Retrieval-Augmented Generation (RAG) represents a significant advancement in addressing the inherent limitations of LLMs. By intelligently incorporating external knowledge bases, RAG effectively reduces model hallucinations and enables access to domain-specific information without requiring costly model retraining. At its core, the RAG architecture consists of two fundamental components:

\begin{itemize}[leftmargin=*]

    \item \textbf{Indexing Module $\varphi(\cdot)$}: This component processes a knowledge database $\mathcal{D}$ (such as document collections in text-based RAG) to create an optimized index structure $\hat{\mathcal{D}} = \varphi(\mathcal{D})$. The data structure enables rapid and efficient knowledge retrieval during query processing. The indexing process transforms raw information into an organized, searchable format facilitating retrieval operations.

    \item \textbf{Retrieval Module $\psi(\cdot)$}: When presented with a user query $q$, this module performs query-specific knowledge retrieval from the indexed data structure, denoted as $\psi(q, \hat{\mathcal{D}})$. The process involves identifying and extracting pertinent information from the indexed knowledge database, yielding informative sources that directly support answering the user's query.
\end{itemize}

In essence, the RAG framework operates in two phases. In the preprocessing phase, the indexing module organizes all data into searchable structures. In the query phase, the retrieval module finds relevant knowledge for each input query $q$. The large language model (LLM) then processes both the query and retrieved knowledge to generate responses, expressed as $\text{LLM}(q, \psi(q, \hat{\mathcal{D}}))$.

\textbf{Retrieval-Augmented Generation with Videos.} While text-based RAG techniques are well-established, their extension to video knowledge remains largely unexplored. Our work advances the capability of Large Language Models (LLMs) to comprehend extremely long videos as a rich knowledge source. We achieve this by: i) Effectively capturing multi-modal characteristics (visual, audio, textual) and their temporal dynamics; ii) Modeling complex cross-modal alignment and inter-dependencies between different information streams.

We formulate this real-world challenge as a novel retrieval scenario with an unconstrained video knowledge base $\mathcal{D} = {\mathcal{V}_1, \ldots, \mathcal{V}_n}$, where each video $\mathcal{V}_i$ can be of arbitrary duration and the total number of videos $n$ is unrestricted. To address this challenge, our \model\ framework enables effective video knowledge discovery and semantic understanding while ensuring comprehensive responses through effectively multi-modal context modeling.
\section{The \model\ Framework}

\begin{figure*}[t]
    \centering
    \includegraphics[width=1.0\textwidth]{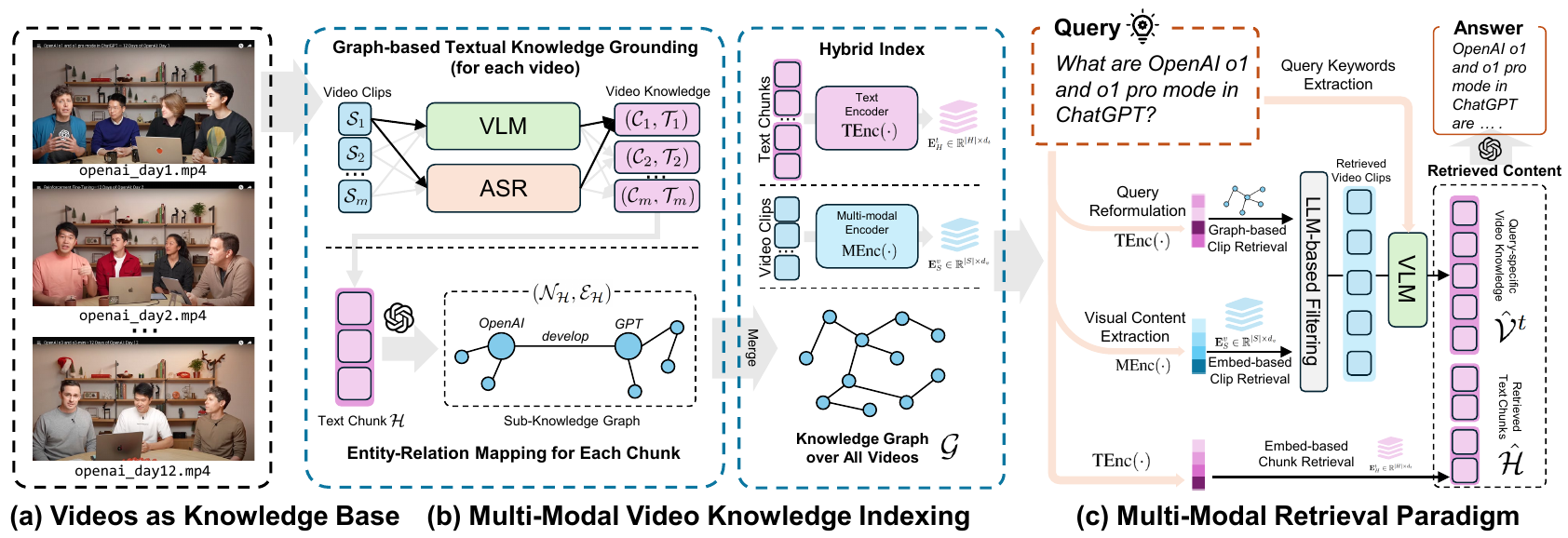}
    \vspace{-0.2in}
    \caption{The overall framework of our proposed RAG framework \model\ for videos.}
    \vspace{-0.1in}
    \label{fig:framework}
\end{figure*}

We present our retrieval-augmented generation framework designed for understanding unlimited-length video content. Our approach addresses two fundamental challenges: (1) multi-modal knowledge indexing that effectively captures and organizes visual, audio, and semantic information from videos, and (2) knowledge-grounded information retrieval that enables precise retrieval of relevant video clips for generating accurate responses through large language models. In the following sections, we detail these components and their integration into a unified video understanding system.

\subsection{Multi-Modal Video Knowledge Indexing}

Unlike traditional text documents, videos encapsulate information through multiple modalities - primarily visual frames - creating unique challenges for knowledge extraction and organization. Standard text-based RAG methods prove insufficient for video content due to several fundamental limitations: (1) text-based systems cannot directly capture visual dynamics; (2) temporal dependencies that traditional RAG approaches fail to preserve across video frames; (3) cross-modal interactions that simple text encoding cannot capture between visual and textual information. 

To address these challenges, we introduce a comprehensive indexing framework with two components: \textbf{Graph-based Textual Knowledge Grounding} that transforms multi-modal signals into structured textual representations while preserving semantic relationships and temporal dependencies, and \textbf{Multi-Modal Context Encoding} that captures fine-grained cross-modal interactions through unified embeddings. This dual-channel architecture enables \model\ to effectively organize and index long-context videos while preserving the semantic richness of multi-modal content.

\subsubsection{\textbf{Graph-based Textual Knowledge Grounding}}
\label{sec:textual indexing}

Our framework transforms multi-modal video content into structured textual knowledge through graph-based techniques for enhanced indexing and retrieval. The conversion process operates across two key modalities: for visual content, we employ state-of-the-art Vision Language Models (VLMs) to generate comprehensive textual descriptions capturing scene dynamics and contextual information; for auditory streams, we leverage high-fidelity Automatic Speech Recognition (ASR) to extract spoken content with temporal alignment. This dual-stream processing ensures both visual semantics and audio information are preserved in our textual knowledge representation.

\begin{itemize}[leftmargin=*]
    \item \textbf{Vision-Text Grounding}: To analyze visual content effectively, we segment each video $\mathcal{V}$ into short clips ${\mathcal{S}_1, \ldots, \mathcal{S}_m}$, enabling processing of unlimited-length videos. For each clip $\mathcal{S}_j$, we transform visual information into text through a two-step process: first, we uniformly sample $k$ frames ($k \leq 10$) chronologically to capture key visual elements; then, we employ Vision-Language Models (VLMs) to generate comprehensive natural language descriptions capturing objects, actions, and scene dynamics. The caption generation process follows:
    \begin{align}
        \mathcal{C}_j = \text{VLM}(\mathcal{T}_j, \{\textbf{F}_1, \ldots, \textbf{F}_k\} \mid \textbf{F} \in \mathcal{S}_j),
    \end{align}
    where ${\textbf{F}}$ denotes the chronologically ordered set of $k$ sampled frames from the clip $\mathcal{S}_j$. We maintain $k \leq 10$ to optimize efficiency while preserving temporal coherence. The model integrates both visual frames and clip transcript $\mathcal{T}_j$ as input prompts, enabling the VLM to generate contextually rich captions $\mathcal{C}_j$ that capture both visual dynamics and associated speech content.
    
    \item \textbf{Audio-Text Grounding}: To capture crucial elements like dialogue and narration that enrich video understanding, we employ Automatic Speech Recognition (ASR) technology to transcribe each video clip, where $\mathcal{T}_j = \text{ASR}(\mathcal{S}_j)$ represents the extracted transcript from the clip $\mathcal{S}_j$.
\end{itemize}

For each video clip $\mathcal{S}$, we then create a unified and semantically rich textual representation by systematically merging the generated visual captions and ASR transcriptions $(\mathcal{C}, \mathcal{T})$. For a video $\mathcal{V}$ containing $m$ sequential clips, we formalize the complete knowledge extraction process as:
\begin{align}
    \mathcal{V}^{t} = \{(\mathcal{C}_l, \mathcal{T}_l) \mid l \in [1, m]\}.
\end{align}
At the core of our \model\ lies the challenge of organizing and retrieving multi-video knowledge efficiently. To address this, we propose a graph-based indexing framework that systematically links textual knowledge across different videos. This architecture enables incremental construction of a comprehensive knowledge graph from the extracted textual information, while maintaining semantic relationships and contextual dependencies. The entire process is executed through these essential steps, each designed to optimize multi-modal knowledge representation and retrieval:
\begin{itemize}[leftmargin=*]
    \item \textbf{Semantic Entity Recognition and Relationship Mapping}:
    Our framework leverages Large Language Models (LLMs) to construct a high-quality knowledge graph $\mathcal{G} = (\mathcal{N}, \mathcal{E})$ that comprehensively captures and connects video knowledge. To optimize LLM performance and manage their context window limitations effectively, we implement a strategic processing pipeline: \\\vspace{-0.12in}
    
    $\bullet$ (i) \textbf{Text Segmentation}. The first stage focuses on text segmentation, where we methodically divide video textual descriptions $\mathcal{V}^{t}$ into manageable chunks $\mathcal{H}_i \in \mathcal{V}^{t}$. Each chunk is carefully constructed to contain multiple video clip descriptions while adhering to a predefined length threshold, ensuring optimal processing while maintaining semantic coherence. \\\vspace{-0.12in}

    $\bullet$ (ii) \textbf{Entity-Relation Extraction}. In the entity-relation extraction phase, we process each chunk's caption-transcript pairs through LLMs to identify key entities (represented as nodes $\mathcal{N}$) and extract meaningful relationships (represented as edges $\mathcal{E}$). For instance, given the text ``\textit{GPT-4 utilizes transformer architecture for advanced natural language understanding, while incorporating visual features through ViT's patch-based image encoding}'', the system extracts entities ``\textit{GPT-4}'', ``\textit{transformer architecture}'', and ``\textit{Vision Transformer (ViT)}'', along with relationships ``\textit{GPT-4 utilizes transformer architecture}'' and ``\textit{GPT-4 incorporates ViT's encoding}'' in the domain of LLMs.

    \item \textbf{Incremental Graph Construction and Cross-Video Knowledge Integration}:
    The construction of our comprehensive knowledge graph follows an iterative and systematic approach across multiple video sources. Our framework implements a sophisticated incremental construction process that ensures coherent knowledge integration through several key mechanisms:

    $\bullet$ (i) \textbf{Entity Unification and Merging}. Our cross-video entity unification process systematically identifies and merges semantically equivalent entities across various videos into unified nodes within the knowledge graph $\mathcal{G}$. This unification approach not only maintains a consistent representation of entities but also preserves the rich contextual information derived from diverse video sources. As a result, it effectively creates a cohesive and interconnected knowledge network that enhances the overall understanding and usability of the information contained within the graph. \\\vspace{-0.12in}

    $\bullet$ (ii) \textbf{Dynamic Knowledge Graph Evolution}. As new video content is processed, our knowledge graph undergoes systematic expansion through dual-track evolution: the integration of newly discovered entities and the establishment of previously unobserved relationships. When processing textual chunks from incoming videos, the system not only identifies and incorporates novel entities (e.g., emerging AI architectures or methodologies) but also discovers new semantic connections between existing nodes. This bidirectional growth process simultaneously reinforces established knowledge patterns while accommodating emerging concepts, ensuring the graph maintains both comprehensiveness and adaptability as it scales. \\\vspace{-0.12in}

    $\bullet$ (iii) \textbf{LLM-Powered Semantic Synthesis}. To maintain semantic coherence, we strategically leverage Large Language Models (LLMs) to generate unified entity descriptions by synthesizing information from multiple video clips. This synthesis process ensures each entity maintains a comprehensive yet consistent representation, effectively consolidating knowledge across different video contexts while preserving semantic accuracy throughout the knowledge structure. \\\vspace{-0.12in}

    Formally, we define the construction of our complete knowledge graph as follows:
    \begin{align}
        \mathcal{G} = (\mathcal{N}, \mathcal{E}) = \bigcup_{\mathcal{H} \in \{ \mathcal{V}_{1}^{t}, \ldots, \mathcal{V}_{n}^{t} \}} (\mathcal{N}_{\mathcal{H}}, \mathcal{E}_{\mathcal{H}}),
    \end{align}
    Let $(\mathcal{N}_{\mathcal{H}}, \mathcal{E}_{\mathcal{H}})$ denote the extracted entities and their relations from each text chunk $\mathcal{H}$, split from the video description $\mathcal{V}^{t}$. Through processing of all videos, we construct the complete graph $\mathcal{G}$.

    \item \textbf{Text Chunk Embedding}. For each text chunk $\mathcal{H}$, we encode a text embedding $\mathbf{e}_{\mathcal{H}}^{t} = \text{TEnc}(\mathcal{H})$ using a text encoder $\text{TEnc}(\cdot)$, enabling efficient retrieval of raw chunks. We denote the complete set of chunks as $H$ and represent their collective text embeddings as $\mathbf{E}_{H}^{t} \in \mathbb{R}^{|H| \times d_{t}}$, where $|H|$ represents the total chunk count and $d_{t}$ denotes the embedding dimension. The knowledge graph $\mathcal{G}$ and chunk embeddings $\mathbf{E}_{H}^{t}$ together form the core components of our graph indexing module.
\end{itemize}

\subsubsection{Multi-Modal Context Encoding}
\label{sec:visual indexing}

In vision-to-text grounding, certain visual nuances are inherently lost, such as lighting dynamics and intricate object details that resist accurate textual representation. To preserve these visual elements, we employ a multi-modal encoder $\text{MEnc}(\cdot)$ that transforms video content into retrieval-optimized embeddings. This encoder is capable of mapping both visual content and textual queries into a shared feature space, enabling efficient semantic retrieval. Building upon powerful multi-modal encoding frameworks like CLIP~\citep{CLIP} and ImageBind~\citep{Imgebind}, we formalize our video encoding as:
\begin{align}
    \mathbf{E}_{S}^{v} \in \mathbb{R}^{|S| \times d_{v}} \quad \textit{w.r.t.} \quad \textbf{e}_{\mathcal{S}}^{v} = \text{MEnc}(\mathcal{S}).
\end{align}
In this formulation, each video clip $\mathcal{S}$ is encoded into an embedding, collectively forming $\mathbf{E}_{S}^{v}$. Here, we utilize the capital $S$ to represent the complete clip set, with $|S|$ denoting the total clip count and $d_{v}$ representing the visual embedding dimensionality. Our \model\ framework's indexing module $\varphi(\cdot)$ processes the video knowledge base $\mathcal{D} = {\mathcal{V}_1, \ldots, \mathcal{V}_n}$ to create a hybrid index combining both knowledge graph and multi-modal context embeddings:
\begin{align}
    \hat{\mathcal{D}} = \varphi(\mathcal{D}) = (\mathcal{G}, \mathbf{E}_{H}^{t}, \mathbf{E}_{S}^{v}).
\end{align}

\subsection{Multi-Modal Retrieval Paradigm}

The Multi-Modal Retrieval Paradigm aims to efficiently retrieve relevant knowledge from videos in response to queries by integrating both textual semantic and visual content matching. Leveraging a hybrid indexing framework $\hat{\mathcal{D}}$, this approach identifies informative video clips and generates query-specific descriptions using VLMs, ensuring a comprehensive retrieval process that captures both semantic understanding and visual context for more accurate responses.
\begin{itemize}[leftmargin=*]
    \item \textbf{Textual Semantic Matching}. The textual retrieval process leverages our constructed knowledge graph $\mathcal{G}$, where each entity contains a text description derived from relevant text chunks. The retrieval process consists of four sequential steps: (i) \textbf{Query Reformulation}: In the initial step, we employ LLMs to reformulate the input query into a declarative sentence, optimizing it for subsequent entity matching operations. (ii) \textbf{Entity Matching}: The system then calculates similarity scores between this reformulated query and entity descriptions within the knowledge graph, identifying the most relevant entities along with their associated text chunks. (iii) \textbf{Chunk Selection}: Following entity matching, we apply a GraphRAG~\cite{GraphRAG}-based methodology to sort and identify the most pertinent chunks ${\mathcal{H}}_{q}$ from the retrieved collection. (iv) \textbf{Video Clip Retrieval}: Finally, we extract video clips from the selected chunks, as each chunk contains descriptions of multiple video clips, resulting in our final textual retrieval set ${\mathcal{S}}_{q}^{t}$.    
    
    \item \textbf{Visual Retrieval via Content Embeddings}. Our framework complements textual matching with direct visual retrieval, enabling semantic alignment between queries and video clips. Building upon our established visual indexing framework (Section~\ref{sec:visual indexing}), each video clip is encoded through a multi-modal encoder $\text{MEnc}(\cdot)$ to generate content-based embeddings. The visual retrieval process operates in two stages: (i) \textbf{Scene Information Extraction from Query:} We leverage LLMs to distill the query $q$ into its core visual scene components, creating a focused scene description. For instance: Original question: ``\emph{In the movie, what color is the car that chases the main character through the city?}''; Scene-focused reformulation: ``\emph{An intense urban chase sequence featuring a car pursuing someone through city streets, with buildings and traffic in the background}'' (ii) \textbf{Cross-Modal Feature Alignment:} This scene-centric query reformulation is projected into the same feature space as our visual embeddings using the multi-modal encoder, leveraging its cross-modal capabilities~\cite{CLIP,Imgebind} to align the context from different modalities. We compute similarity scores between the query embedding and video clip embeddings $\mathbf{E}_{S}^{v}$ through cosine similarity, denoted as $\text{Sim}(\text{MEnc}(q), \mathbf{E}_{S}^{v})$. The top-K matching clips form the visual retrieval result ${\mathcal{S}}_{q}^{v}$.
    
    \item \textbf{LLMs-based Video Clip Filtering}: To filter out noisy clips from the retrieved results, we employ LLMs to evaluate each clip $\mathcal{S} \in {\mathcal{S}}_{q}^{t} \cap {\mathcal{S}}_{q}^{v}$ for its relevance to query $q$ using textual and visual information $\mathcal{V}^{t}_{\mathcal{S}}$ (Section~\ref{sec:textual indexing}). The filtered clips are formally defined as:
    \begin{align}
        \{\hat{\mathcal{S}} \mid (\hat{\mathcal{S}} \in \{\mathcal{S}\}_{q}^{t} \cap \{\mathcal{S}\}_{q}^{v}) \land  \text{LLMs-Judge}(\mathcal{V}^{t}_{\hat{\mathcal{S}}}) = 1\},
    \end{align}
    where the function $\text{LLMs-Judge}(\cdot)$ serves as a binary decision maker that evaluates clip relevance via carefully-designed prompting instructions, returning 1 if the clip contains information vital to answering query $q$. This approach leverages LLMs' advanced semantic reasoning capabilities to effectively filter out irrelevant clips while preserving key information.
    
\end{itemize}

\subsection{Query-Aware Content Integration and Response Generation}

With the retrieved video clips, we implement a two-stage content extraction process. First, we utilize LLMs to extract keywords $\mathcal{K}_{q}$ from query $q$, which are then integrated into VLM prompts alongside sampled frames to generate detailed visual captions $\hat{\mathcal{C}}$:
\begin{align}
    \hat{\mathcal{C}} = \text{VLM}(\mathcal{K}_q, \hat{\mathcal{T}}, \{\textbf{F}_1, \ldots, \textbf{F}_{\hat{k}}\} \mid \textbf{F} \in \hat{\mathcal{S}}),
\end{align}
where $\hat{\mathcal{T}}$ represents the audio transcription for clip $\hat{\mathcal{S}}$, with a larger $\hat{k} > k$ sampled frames. For each clip $\hat{\mathcal{S}}_j$, we create a comprehensive description $\hat{\mathcal{V}}_j^{t} = (\hat{\mathcal{C}}_j, \hat{\mathcal{T}}_j)$ by combining its visual caption and transcription. These descriptions are collected into set ${ \hat{\mathcal{V}}^{t} }$ for enhanced generation. We then enrich this visual analysis with traditional text-based retrieval, employing semantic similarity matching between query $q$ and text chunks ${ \mathcal{H} }$ to obtain relevant textual information ${ \hat{\mathcal{H}} }$. The complete output of our retrieval module $\psi(\cdot)$ thus comprises both query-specific video descriptions and relevant text chunks: $\psi(q, \hat{\mathcal{D}}) = ({ \hat{\mathcal{V}}^{t} }, { \hat{\mathcal{H}} })$. Finally, \model\ leverages a general-purpose LLM (\eg, GPT4 or DeepSeek) to generate responses based on the query $q$ and retrieved content, as detailed in Section~\ref{sec:preliminary}.
\section{Evaluation}
We conduct comprehensive empirical evaluations of our \model\ framework on established benchmark datasets to address the following key research questions (RQs): \textbf{RQ1}: How effectively does \model\ perform in handling long-form video content compared to existing RAG alternative approaches? \textbf{RQ2}: What advantages does \model\ demonstrate over large vision models (LVMs) in understanding extremely long-context videos? \textbf{RQ3}: How do ablation studies reveal the effectiveness of individual components (textual and visual retrieval) in \model? \textbf{RQ4}: What insights can be derived from qualitative case studies of \model\ across diverse application scenarios?

\subsection{Experimental Settings}

\begin{table}[t]
\centering
\caption{Statistics of the experimental dataset \textit{LongerVideos}.}
\label{tab:stats}
\resizebox{0.9\textwidth}{!}{
\begin{tabular}{@{}l|cccccc@{}}
\toprule
Video Type & \#video list & \#video & \#query & \#avg. queries per list & \#overall duration \\
\midrule
\textbf{Lecture}        & 12 & 135 & 376 & 31.3 & $\sim$~64.3 hours\\ 
\textbf{Documentary}    &  5 &  12 & 114 & 22.8 & $\sim$~28.5 hours\\
\textbf{Entertainment}  &  5 &  17 & 112 & 22.4 & $\sim$~41.9 hours\\
\midrule
\textbf{All}            & 22 & 164 & 602 & 27.4 & $\sim$~134.6 hours\\
\bottomrule
\end{tabular}
}
\end{table}

\textbf{Evaluation Datasets.} Current benchmarks for video-based question answering are limited by relatively short durations (average <1 hour per video)~\cite{Video-MME} or single-video understanding scenarios (e.g., MLVU~\cite{MLVU} and LVBench\cite{LVBench}). These constraints make it challenging to evaluate models' capabilities in processing and reasoning across multiple extremely long-context videos for question-answering. To address this limitation in existing evaluation frameworks, we introduce \textit{\textbf{LongerVideos}}, a comprehensive benchmark comprising over twenty video collections across three distinct categories:
\begin{itemize}[leftmargin=*]
    \item \textbf{Lecture Video}: Open-access educational content featuring contemporary technical topics, including AI Agents and RAG Techniques, delivered through comprehensive tutorials.
    \item \textbf{Documentary Video}: High-quality documentaries spanning wildlife exploration, natural landscapes, and expert interviews, each produced with professional cinematography.
    \item \textbf{Entertainment Video}: Diverse content including award ceremonies, gaming commentary with strategic analysis, and travel experiences documenting global cultural explorations.
\end{itemize}
All content is sourced from open-access YouTube videos, ensuring broad accessibility and reproducibility. Using NotebookLM\footnote{\url{https://notebooklm.google/}}, we systematically generate an average of 25+ high-quality queries per collection by processing video transcripts. Each collection averages over 4 hours in total duration, containing between 1 to 20+ individual videos, ultimately yielding a robust evaluation set of 600+ diverse queries. Detailed statistical analysis of the benchmark is presented in Table~\ref{tab:stats}.

\textbf{Evaluation Protocols and Metrics.} We implement two distinct protocols to evaluate model performance across different scenarios. The first protocol, \textbf{Win-Rate Comparison}, follows established Retrieval-Augmented Generation (RAG) evaluation methodologies~\cite{GraphRAG, LightRAG} using LLM-based judgment. This approach employs \texttt{GPT-4o-mini} to comparatively rank responses generated by two models, providing explanatory justification for each ranking decision. The second protocol, \textbf{Quantitative Comparison}, extends the LLM-based judgment by incorporating score assignment. It establishes a standard baseline answer for each query, against which other responses are evaluated on a 5-point scale, ranging from 1 (strongly worse) to 5 (strongly better).

We strategically apply these protocols for different evaluation purposes. The Win-Rate Comparison protocol is utilized to assess our methods against various RAG techniques and their ablation variants, enabling competitive analysis of our \model. Conversely, the Quantitative Comparison protocol facilitates fine-grained analysis when comparing \model\ with long video understanding methods. Following the framework established in~\cite{GraphRAG}, our evaluation encompasses multiple dimensions for comprehensive analysis, focusing on five distinct aspects detailed as follows:

(i) \textit{\textbf{Comprehensiveness}} evaluates answer coverage of question aspects. (ii) \textit{\textbf{Empowerment}} measures how effectively the answer enables reader understanding and informed judgment. (iii) \textit{\textbf{Trustworthiness}} assesses the answer's credibility through detail sufficiency and alignment with common knowledge. (iv) \textit{\textbf{Depth}} examines the presence of thorough analysis versus superficial information. (v) \textit{\textbf{Density}} evaluates the concentration of relevant information while minimizing redundancy.

We implement two key strategies to ensure reliable results. First, to mitigate position-related bias in LLM inference, we alternate the answer positions within each prompt and collect two judgments per query during win-rate comparisons. Second, to minimize statistical variance, we perform five evaluation repetitions for both win-rate and quantitative assessments, then aggregate wins or calculate mean scores to determine final results. Complete evaluation prompts are provided in Appendix~\ref{apd:prompt4evaluation}.

\textbf{Implementation Details of \model}. For vision-text grounding (Section~\ref{sec:textual indexing}), we segment videos into 30-second clips and use $k=5$ frames for initial visual captioning. We employ the quantized MiniCPM-V~\cite{MiniCPMV} as the VLM model and Distil-Whisper~\cite{Whisper, DistilWhisper} as the VSR model. For multi-modal encoding (Section~\ref{sec:visual indexing}), we utilize ImageBind~\cite{Imgebind} as $\text{MEnc}(\cdot)$ for both visual and textual encoding. Entity and textual chunk retrieval leverage OpenAI's \texttt{text-embedding-3-small} model, while section-run visual captioning uses an increased frame count of $\hat{k}=15$. Throughout the implementation, \texttt{GPT-4o-mini} serves as our core LLM for indexing, retrieval, and answer generation. Complete implementation details are available in our open-source codebase.

\subsection{Overall Comparison (RQ1 \& RQ2)}

\begin{table}[t]
\centering
\caption{We analyze the performance of \model\ against RAG baselines on the LongerVideos dataset, presenting results both by individual video categories and across the complete dataset.}
\label{tab:rag performance}
\resizebox{\textwidth}{!}{
\begin{tabular}{@{}lcccccccc@{}}
\toprule
\textbf{}    & \multicolumn{2}{c}{\textbf{Lecture}} & \multicolumn{2}{c}{\textbf{Documentary}} & \multicolumn{2}{c}{\textbf{Entertainment}} & \multicolumn{2}{c}{\textbf{All}} \\ 
\cmidrule(lr){2-3} \cmidrule(lr){4-5} \cmidrule(lr){6-7} \cmidrule(lr){8-9}
                      & NaiveRAG & \textbf{\model} & NaiveRAG & \textbf{\model} & NaiveRAG & \textbf{\model} & NaiveRAG & \textbf{\model} \\
\midrule
Comprehensiveness      & 47.63\%      & \underline{52.37}\%     & 44.04\%      & \underline{55.96}\%     & 46.43\%      & \underline{53.57}\%     & 46.73\%      & \underline{53.27}\%     \\
Empowerment            & 45.85\%      & \underline{54.15}\%     & 40.00\%      & \underline{60.00}\%     & 45.36\%      & \underline{54.64}\%     & 44.65\%      & \underline{55.35}\%     \\
Trustworthiness        & 46.73\%      & \underline{53.27}\%     & 42.54\%      & \underline{57.46}\%     & 44.46\%      & \underline{55.54}\%     & 45.51\%      & \underline{54.49}\%     \\
Depth                  & 46.70\%      & \underline{53.30}\%     & 43.25\%      & \underline{56.75}\%     & 46.07\%      & \underline{53.93}\%     & 45.93\%      & \underline{54.07}\%     \\
Density                & 46.73\%      & \underline{53.27}\%     & 44.21\%      & \underline{55.79}\%     & 44.29\%      & \underline{55.71}\%     & 45.80\%      & \underline{54.20}\%     \\
Overall Winner         & 47.66\%      & \underline{52.34}\%     & 44.04\%      & \underline{55.96}\%     & 46.43\%      & \underline{53.57}\%     & 46.74\%      & \underline{53.26}\%     \\
\cmidrule(lr){2-3} \cmidrule(lr){4-5} \cmidrule(lr){6-7} \cmidrule(lr){8-9}
                      & GraphRAG-$l$ & \textbf{\model} & GraphRAG-$l$ & \textbf{\model} & GraphRAG-$l$ & \textbf{\model} & GraphRAG-$l$ & \textbf{\model} \\
\midrule
Comprehensiveness      & 44.60\%      & \underline{55.40}\%     & 48.68\%      & \underline{51.32}\%     & 49.29\%      & \underline{50.71}\%     & 46.25\%      & \underline{53.75}\%     \\
Empowerment            & 42.34\%      & \underline{57.66}\%     & 47.54\%      & \underline{52.46}\%     & 49.02\%      & \underline{50.98}\%     & 44.57\%      & \underline{55.43}\%     \\
Trustworthiness        & 42.79\%      & \underline{57.21}\%     & 47.11\%      & \underline{52.89}\%     & 46.07\%      & \underline{53.93}\%     & 44.22\%      & \underline{55.78}\%     \\
Depth                  & 42.34\%      & \underline{57.66}\%     & 48.33\%      & \underline{51.67}\%     & 49.55\%      & \underline{50.45}\%     & 44.82\%      & \underline{55.18}\%     \\
Density                & 39.26\%      & \underline{60.74}\%     & 45.26\%      & \underline{54.74}\%     & 46.52\%      & \underline{53.48}\%     & 41.74\%      & \underline{58.26}\%     \\
Overall Winner         & 44.44\%      & \underline{55.56}\%     & 48.68\%      & \underline{51.32}\%     & 49.20\%      & \underline{50.80}\%     & 46.13\%      & \underline{53.87}\%     \\
\cmidrule(lr){2-3} \cmidrule(lr){4-5} \cmidrule(lr){6-7} \cmidrule(lr){8-9}
                      & GraphRAG-$g$ & \textbf{\model} & GraphRAG-$g$ & \textbf{\model} & GraphRAG-$g$ & \textbf{\model} & GraphRAG-$g$ & \textbf{\model} \\
\midrule
Comprehensiveness      & 42.66\%      & \underline{57.34}\%     & 46.23\%      & \underline{53.77}\%     & 48.48\%      & \underline{51.52}\%     & 44.42\%      & \underline{55.58}\%     \\
Empowerment            & 39.55\%      & \underline{60.45}\%     & 44.04\%      & \underline{55.96}\%     & 48.30\%      & \underline{51.70}\%     & 42.03\%      & \underline{57.97}\%     \\
Trustworthiness        & 38.54\%      & \underline{61.46}\%     & 41.49\%      & \underline{58.51}\%     & 43.48\%      & \underline{56.52}\%     & 40.02\%      & \underline{59.98}\%     \\
Depth                  & 40.61\%      & \underline{59.39}\%     & 45.26\%      & \underline{54.74}\%     & 47.23\%      & \underline{52.77}\%     & 42.72\%      & \underline{57.28}\%     \\
Density                & 37.55\%      & \underline{62.45}\%     & 46.93\%      & \underline{53.07}\%     & 48.04\%      & \underline{51.96}\%     & 41.28\%      & \underline{58.72}\%     \\
Overall Winner         & 42.23\%      & \underline{57.77}\%     & 46.32\%      & \underline{53.68}\%     & 48.75\%      & \underline{51.25}\%     & 44.22\%      & \underline{55.78}\%     \\
\cmidrule(lr){2-3} \cmidrule(lr){4-5} \cmidrule(lr){6-7} \cmidrule(lr){8-9}
                      & LightRAG & \textbf{\model} & LightRAG & \textbf{\model} & LightRAG & \textbf{\model} & LightRAG & \textbf{\model} \\
\midrule
Comprehensiveness      & 42.42\%      & \underline{57.58}\%     & 45.09\%      & \underline{54.91}\%     & 43.84\%      & \underline{56.16}\%     & 43.19\%      & \underline{56.81}\%     \\
Empowerment            & 39.55\%      & \underline{60.45}\%     & 38.95\%      & \underline{61.05}\%     & 42.05\%      & \underline{57.95}\%     & 39.90\%      & \underline{60.10}\%     \\
Trustworthiness        & 39.52\%      & \underline{60.48}\%     & 42.11\%      & \underline{57.89}\%     & 40.00\%      & \underline{60.00}\%     & 40.10\%      & \underline{59.90}\%     \\
Depth                  & 40.13\%      & \underline{59.87}\%     & 41.93\%      & \underline{58.07}\%     & 41.96\%      & \underline{58.04}\%     & 40.81\%      & \underline{59.19}\%     \\
Density                & 39.57\%      & \underline{60.43}\%     & 42.37\%      & \underline{57.63}\%     & 41.61\%      & \underline{58.39}\%     & 40.48\%      & \underline{59.52}\%     \\
Overall Winner         & 42.15\%      & \underline{57.85}\%     & 44.30\%      & \underline{55.70}\%     & 43.75\%      & \underline{56.25}\%     & 42.86\%      & \underline{57.14}\%     \\
\bottomrule
\end{tabular}
}
\end{table}

We assess \model's capabilities in comprehending long-form, multi-video content by comparing its retrieval-augmented generation performance against state-of-the-art RAG baselines.
\begin{itemize}[leftmargin=*]
    \item \textbf{NaiveRAG~\cite{NaiveRAG}}: A standard RAG implementation that segments documents into uniform-sized chunks and retrieves contextually relevant content through text embedding similarity matching, serving as a widely-adopted baseline for retrieval-augmented generation systems.
    \item \textbf{GraphRAG~\cite{GraphRAG}}: An enhanced RAG system that that leverages LLMs to construct entity knowledge graphs from input documents. It improves answer generation by performing community-based graph summarization to capture global context and relationships between entities.
    \item \textbf{LightRAG~\cite{LightRAG}}: A lightweight graph-based RAG framework that implements dual-level retrieval architecture, integrating both low-level and high-level semantic knowledge discovery. The system enables efficient and contextually-aware document retrieval to process complex queries.
\end{itemize}

\textbf{Details of Baseline Implementation.} To ensure fair comparison, we implement all baseline methods with the following specifications: $\bullet$ \textbf{Input Data}: We utilize grounded textual knowledge (\eg, visual captions and transcripts) from all videos, employing identical chunk-splitting protocols as our method. $\bullet$ \textbf{Visual Processing}: For frame-level analysis, we maintain 15 frames per video clip for visual caption generation, matching the fine-grained section-run captions produced by \model's retrieval process. $\bullet$ \textbf{Baseline Variants}: GraphRAG: Implemented with both local (GraphRAG-$l$) and global (GraphRAG-$g$) search capabilities; LightRAG: Deployed with full hybrid search functionality.

\textbf{Comparison Results and Analysis (RQ1)}.
Table~\ref{tab:rag performance} presents the win rate evaluation results comparing \model\ with baseline methods. Our analysis reveals several significant findings:

\begin{itemize}[leftmargin=*]

    \item \textbf{Superior Video-based RAG Performance}. Our evaluation demonstrates that \model\ consistently outperforms all baseline methods across performance metrics. The superior performance stems from our innovative multi-modal video knowledge indexing framework, which combines graph-based knowledge grounding with multi-modal context encoding, enabling effective capture and organization of visual dynamics and semantic information across videos.\\\vspace{-0.12in}

    Furthermore, \model's multi-modal retrieval paradigm significantly enhances performance through its hybrid approach to knowledge discovery. By integrating textual semantic matching with visual content embedding-based retrieval, the system achieves precise and contextually relevant video clip retrieval. This comprehensive retrieval strategy enables more accurate and nuanced responses compared to traditional single-modality approaches, while ensuring the extracted information remains semantically coherent and contextually appropriate. \\\vspace{-0.12in}

    \item \textbf{Performance Analysis Across Baseline Methods}. In comparison with NaiveRAG, \model\ demonstrates exceptional performance across evaluation dimensions, with particular strengths in Comprehensiveness and Empowerment. This superiority stems from our effective knowledge indexing framework, which interlinks information across multiple videos, enabling sophisticated synthesis of diverse information during retrieval and yielding more comprehensive responses. \\\vspace{-0.12in}

    When benchmarked against GraphRAG and LightRAG, \model\ achieves superior performance through multi-modal context integration capabilities. Our approach excels in two aspects: (1) sophisticated knowledge indexing that effectively fuses visual-textual information, and (2) query-aware retrieval that leverages unified multi-modal representations for precise content selection. This architecture enables more nuanced understanding and contextually coherent response generation, significantly outperforming existing methods in knowledge-grounded video question answering.
    
\end{itemize}

\begin{table}[t]
\centering
\caption{We conduct quantitative comparisons between \model\ and existing long-context video understanding models on the benchmark dataset. Each model's performance is rated against NaiveRAG (our baseline) on a 5-point scale, where 1 indicates 'strongly worse than baseline' and 5 represents 'strongly better than baseline'. We evaluate across three video categories: lectures ('lec'), documentaries ('doc'), and entertainment content ('ent'), with 'all' representing the aggregate performance.}
\label{tab:vision performance}
\resizebox{\textwidth}{!}{
\begin{tabular}{@{}lcccccccccccccccc@{}}
\toprule
\textbf{}    & \multicolumn{4}{c}{LLaMA-VID} & \multicolumn{4}{c}{VideoAgent} & \multicolumn{4}{c}{NotebookLM} & \multicolumn{4}{c}{\textbf{\model}} \\ 
\cmidrule(lr){2-5} \cmidrule(lr){6-9} \cmidrule(lr){10-13} \cmidrule(lr){14-17}
& lec & doc & ent & \textbf{all} & lec & doc & ent & \textbf{all} & lec & doc & ent & \textbf{all} & lec & doc & ent & \textbf{all} \\
\midrule
Comprehensiveness      & 2.36 & 2.62 & 2.54 & 2.44 & 2.02 & 1.99 & 1.80 & 1.98 & 3.53 & 3.20 & 2.96 & 3.36 & 4.48 & 4.51 & 4.44 & \underline{4.48} \\
Empowerment            & 2.79 & 3.03 & 2.86 & 2.85 & 2.42 & 2.37 & 2.10 & 2.35 & 3.88 & 3.62 & 3.29 & 3.72 & 4.51 & 4.55 & 4.45 & \underline{4.51} \\
Trustworthiness        & 3.15 & 3.30 & 3.35 & 3.22 & 2.83 & 2.73 & 2.65 & 2.78 & 3.95 & 3.80 & 3.71 & 3.88 & 4.50 & 4.54 & 4.48 & \underline{4.50} \\
Depth                  & 2.01 & 2.06 & 2.00 & 2.02 & 1.79 & 1.75 & 1.62 & 1.75 & 3.14 & 2.89 & 2.55 & 2.98 & 4.34 & 4.42 & 4.31 & \underline{4.35} \\
Density                & 3.15 & 3.28 & 3.21 & 3.18 & 2.82 & 2.73 & 2.52 & 2.75 & 4.07 & 3.82 & 3.61 & 3.94 & 4.59 & 4.63 & 4.56 & \underline{4.59} \\
Overall Score          & 2.36 & 2.61 & 2.54 & 2.44 & 2.03 & 2.01 & 1.80 & 1.98 & 3.54 & 3.21 & 2.97 & 3.37 & 4.45 & 4.49 & 4.41 & \underline{4.45} \\

\bottomrule
\end{tabular}
}
\end{table}

To establish comprehensive performance benchmarks, we evaluate \model\ against state-of-the-art large vision models specifically designed for long-context video understanding, encompassing both advanced vision-language models and intelligent agent systems.
\begin{itemize}[leftmargin=*]
    
    \item \textbf{LLaMA-VID~\cite{Llama-VID}}: A vision-language framework that leverages context and content tokens for efficient long video processing, addressing token complexity in video understanding tasks.
    
    \item \textbf{VideoAgent~\cite{VideoAgent}}: A multi-modal agent that integrates diverse foundation models through a unified memory architecture. It enables powerful video understanding through fine-grained object detection, tracking, and modeling of temporal dependencies within short video clips.
    
    \item \textbf{NotebookLM}: An assistant system by Google designed for video content analysis. It enables efficient multi-video comprehension and information retrieval through advanced transcript analysis, allowing users to extract contextually coherent insights across multiple video sources.
    
\end{itemize}

\textbf{Details of Baseline Implementation}. We implement all baselines using their official codebases or available platforms for fair evaluation. For vision-language model baselines like LLaMA-VID, we standardize the implementation through three crucial modifications: (i) Replacing their original ASR model with the one used in \model\ for consistent transcript extraction; (ii) Uniformly sampling 3,600 frames per video due to GPU memory constraints (48GB per GPU); and (iii) Employing prompting instructions consistent with the compared RAG methods.

\textbf{Comparison Results and Analysis (RQ2)}. We present comprehensive performance comparisons between \model\ and existing video understanding methods in Table~\ref{tab:vision performance}. The results demonstrate that our model consistently outperforms all compared long-context video understanding methods across various metrics. We attribute these improvements to the following aspects:

\begin{itemize}[leftmargin=*]
    \item \textbf{Enhanced Long-Context Modeling}. \model\ proposes a graph-enhanced multi-modal indexing and retrieval mechanism that significantly extends video processing capabilities beyond existing vision models. Unlike conventional approaches that face length constraints, our model can effectively handle unlimited long-context videos by establishing and leveraging cross-video knowledge connections and inter-dependent relations. This architectural advantage enables comprehensive knowledge extraction and integration across extended video sequences, surpassing models like LLaMA-VID that are limited by computational constraints when processing video frames directly.
    
    \item \textbf{Superior Multi-Modal Fusion}. \model\ excels in capturing, reasoning, and aligning diverse multi-modal contexts through advanced cross-modal knowledge integration. Our approach effectively fuses information across different modalities (visual, audio, and textual), enabling superior cross-modal alignment and comprehensive understanding. This multi-modal synthesis surpasses single-modality focused approaches like VideoAgent (visual-only) and NotebookLM (transcript-only), leading to more nuanced, coherent, and expressive video understanding.
\end{itemize}

\subsection{Ablation Study (RQ3)} 
To evaluate the effectiveness of our multi-modal indexing and retrieval design, we conduct comprehensive ablation studies using two model variants: $\bullet$ \textbf{Variant 1 (-\textit{Graph}):} Removes the graph-based index-retrieval pipeline, limiting the model's ability to establish multi-video relationships. $\bullet$ \textbf{Variant 2 (-\textit{Vision}):} Eliminates the visual indexing and retrieval component from the multi-modal encoder.

\begin{figure*}[t]
    \centering
    \includegraphics[width=\textwidth]{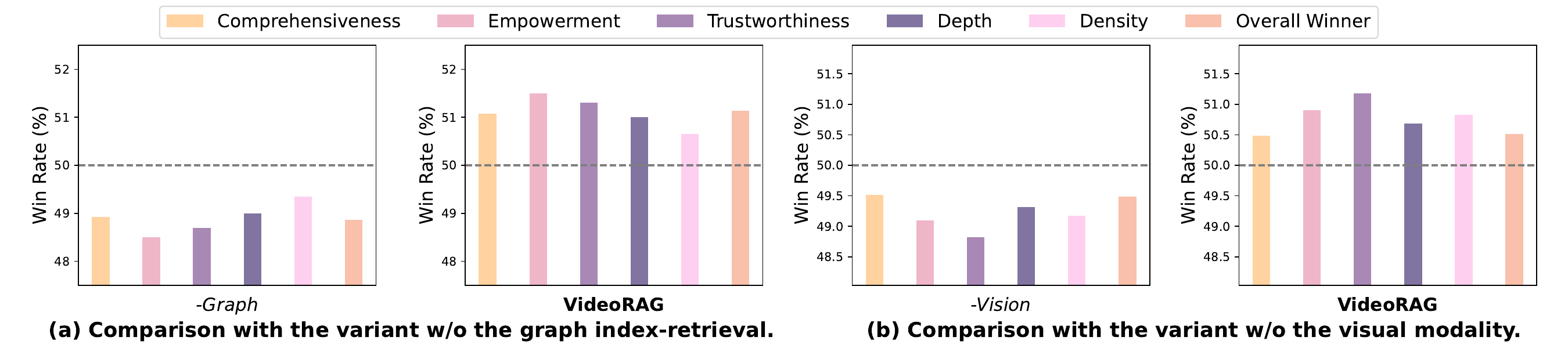}
    \caption{Ablation on graph-based knowledge grounding and cross-modal retrieval components.}
    \vspace{-0.2in}
    \label{fig:ablation study}
\end{figure*}

The ablation study results in Figure~\ref{fig:ablation study} reveal the crucial contribution of each component to the model performance of \model, as evidenced by the following analyses:

\begin{itemize}[leftmargin=*]
\item The -\textit{Graph} variant exhibits significant performance degradation across all evaluation metrics, demonstrating that our graph-based index-retrieval mechanism is essential for two key capabilities: (1) capturing complex inter-video relationships and (2) establishing cross-video knowledge dependencies. This validates the effectiveness of our graph-enhanced architecture in connecting and synthesizing information across multiple videos.

\item The -\textit{Vision} variant shows substantially decreased win rates, underscoring the critical role of visual information processing in our framework. This performance drop validates our model's effective multi-modal context fusion mechanism, which successfully integrates and aligns visual features with other modalities for comprehensive video understanding.

\end{itemize}

These comprehensive findings underscore that the synergistic integration of graph-based architecture and visual modality processing serves as a cornerstone for achieving superior performance in multi-modal indexing and retrieval tasks, validating our architectural design choices.

\subsection{Case Study Analysis (RQ4)}
\label{sec:case study}
\begin{table*}[t]
\centering
\caption{A case study examining \model's response to a query about ``\emph{graders}'' in OpenAI's 12-day edu    cational series demonstrates its video retrieval capabilities and knowledge synthesis.}
\label{tab:case study}
\small
\begin{tabularx}{\textwidth}{X X}

\toprule
\textbf{Query:} Explain the purpose and functionality of 'graders' in the context of reinforcement fine-tuning.\\

\midrule
\textbf{Input Videos:} \href{https://www.youtube.com/playlist?list=PLOXw6I10VTv9lin5AzsHAHCTrC7BdVdEM}{12 Days of OpenAI} (A total of 12 videos) \\

\midrule

\textbf{\model\ Response:} 

\texttt{\#\#\# Understanding Gradation in Reinforcement Fine-Tuning}

\texttt{**Purpose of Graders**}

\texttt{Graders are integral components in the process of reinforcement fine-tuning, mainly utilized to evaluate and improve the performance of machine learning models...} (Omitted for brevity)

\texttt{**Functionality of Graders**}

\texttt{\colorbox{lightgreen}{The grading system operates by comparing a model's outputs against those of}}

\texttt{\colorbox{lightgreen}{verified correct responses. The mechanics of this assessment involve assigning}}

\texttt{\colorbox{lightgreen}{scores to the model's outputs based on their correctness. These scores range}}

\texttt{\colorbox{lightgreen}{from zero to one, where:}}

\texttt{\colorbox{lightblue}{- A score of **0** indicates that the model's answer is completely incorrect.}}

\texttt{\colorbox{lightblue}{- A score of **1** signals that the answer is fully correct.}}

\texttt{- Scores can also be fractional, allowing for partial credit. For instance, \colorbox{lightpink}{a score of **0.7** may denote that the model's suggestion was }}
\texttt{\colorbox{lightpink}{correct but not the highest in a ranked list of possible answers.}} 

\\

\texttt{The process includes:}

\texttt{1. **Input Comparison**: The grader takes both the model’s output and the correct answer, analyzing them to derive a score.}

\texttt{2. **Graded Outputs**: This interaction allows for a dynamic reflection of the model's performance, where the scores directly influence how the model is fine-tuned in subsequent training iterations.}

\texttt{3. **Feedback Loop**: The scores obtained from graders provide feedback that can be used to adjust learning parameters and improve the model’s future responses. This highlights the feedback mechanism that is crucial for reinforcement learning.}

\\

(The remaining content are omitted for brevity.)

\\

\midrule
\textbf{Retrieved Sources by \model:}

\model\ accurately retrieved four continuous video segments from the OpenAI's 12-day show, ranging from 10:00 to 12:00 in Day 2 of the video "\href{https://www.youtube.com/watch?v=yCIYS9fx56U}{Reinforcement Fine-Tuning}." Here, we highlight key moments relevant to the detailed content in the answer. From left to right, these are retrieved moments at timestamps \colorbox{lightgreen}{10:35}, \colorbox{lightblue}{10:39}, and \colorbox{lightpink}{11:10}, which provide informative insights that help \model\ give a comprehensive answer to the query.

\\

\begin{tabular}{ccc}
    {\includegraphics[width=0.3\textwidth]{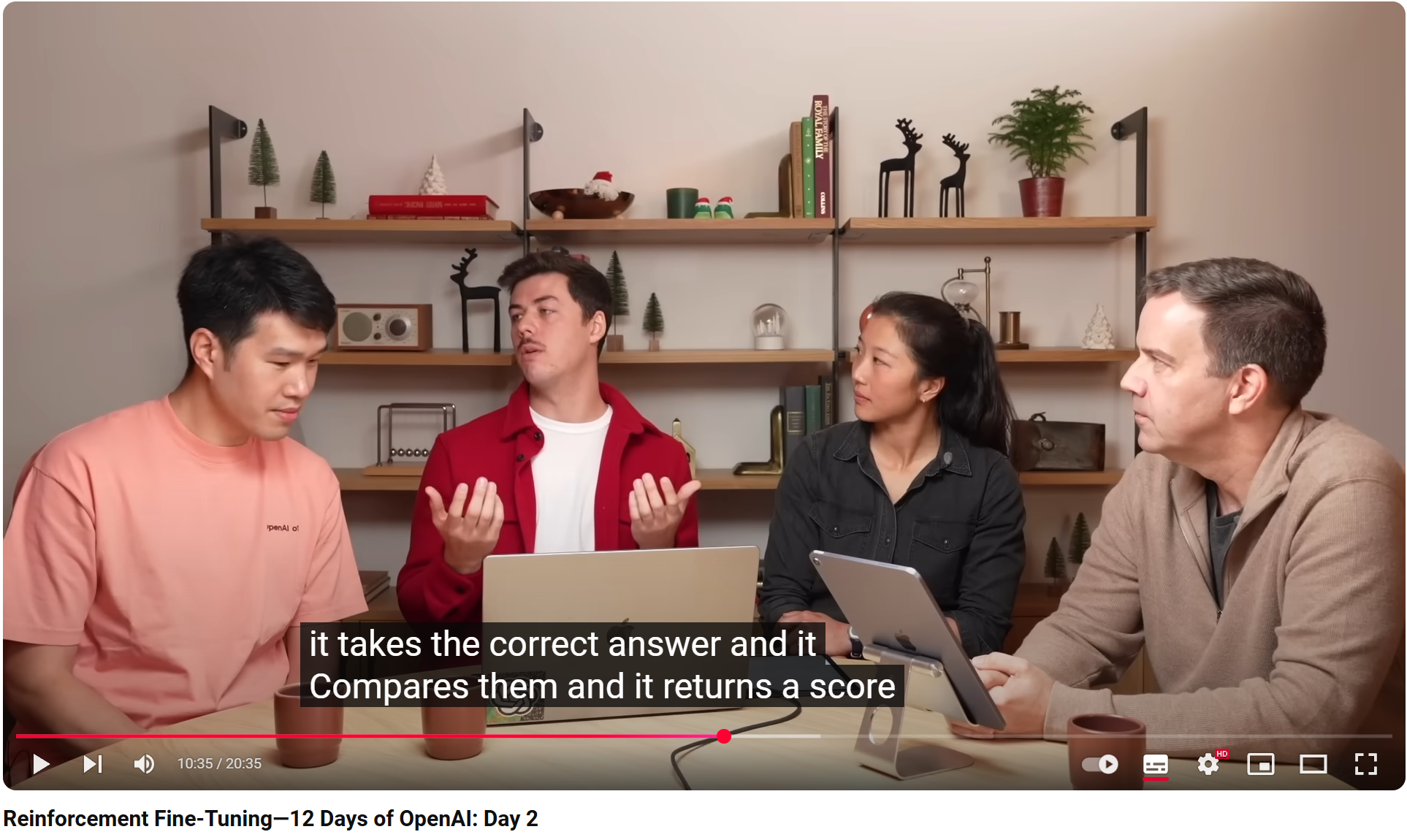}} &
    {\includegraphics[width=0.3\textwidth]{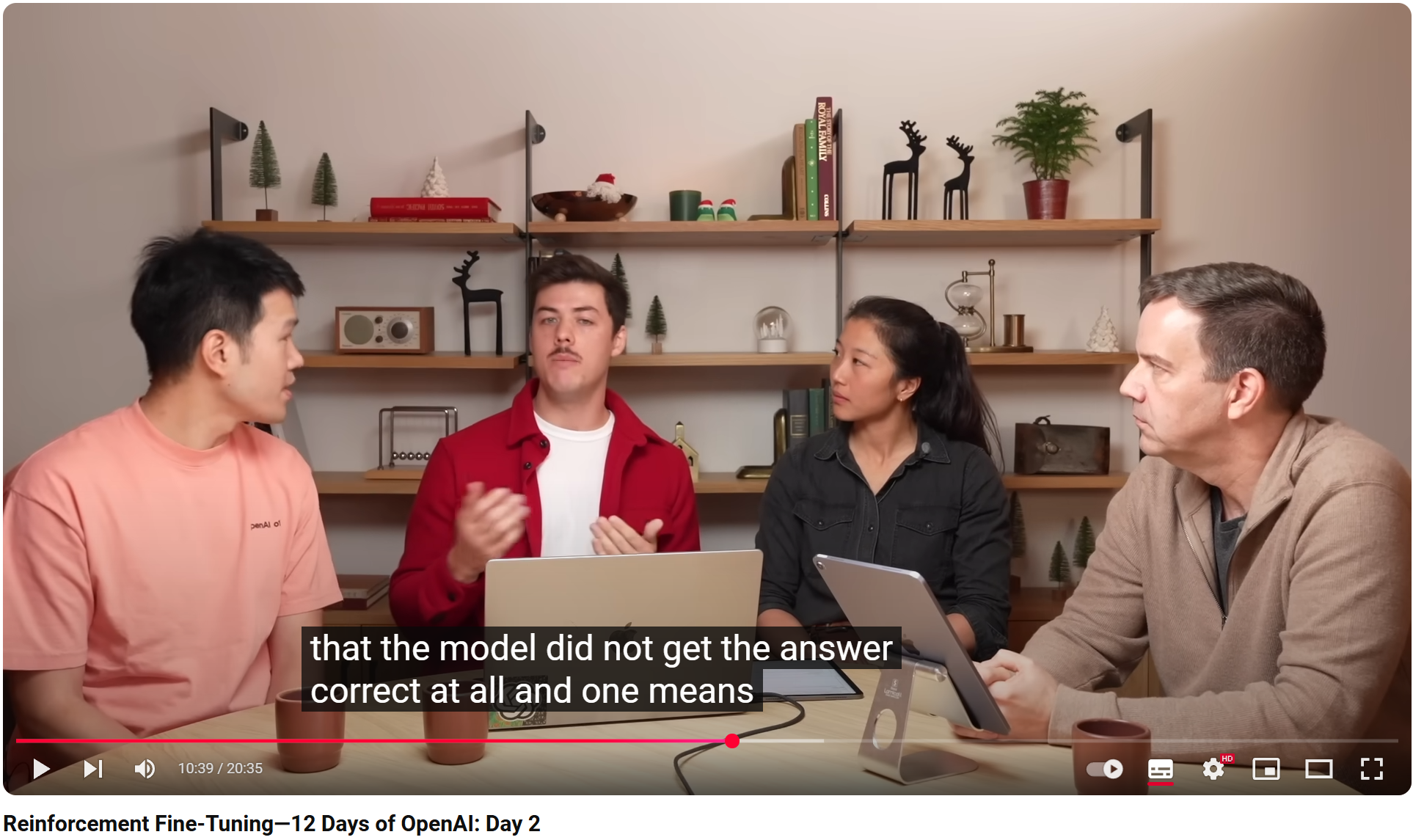}} &
    {\includegraphics[width=0.3\textwidth]{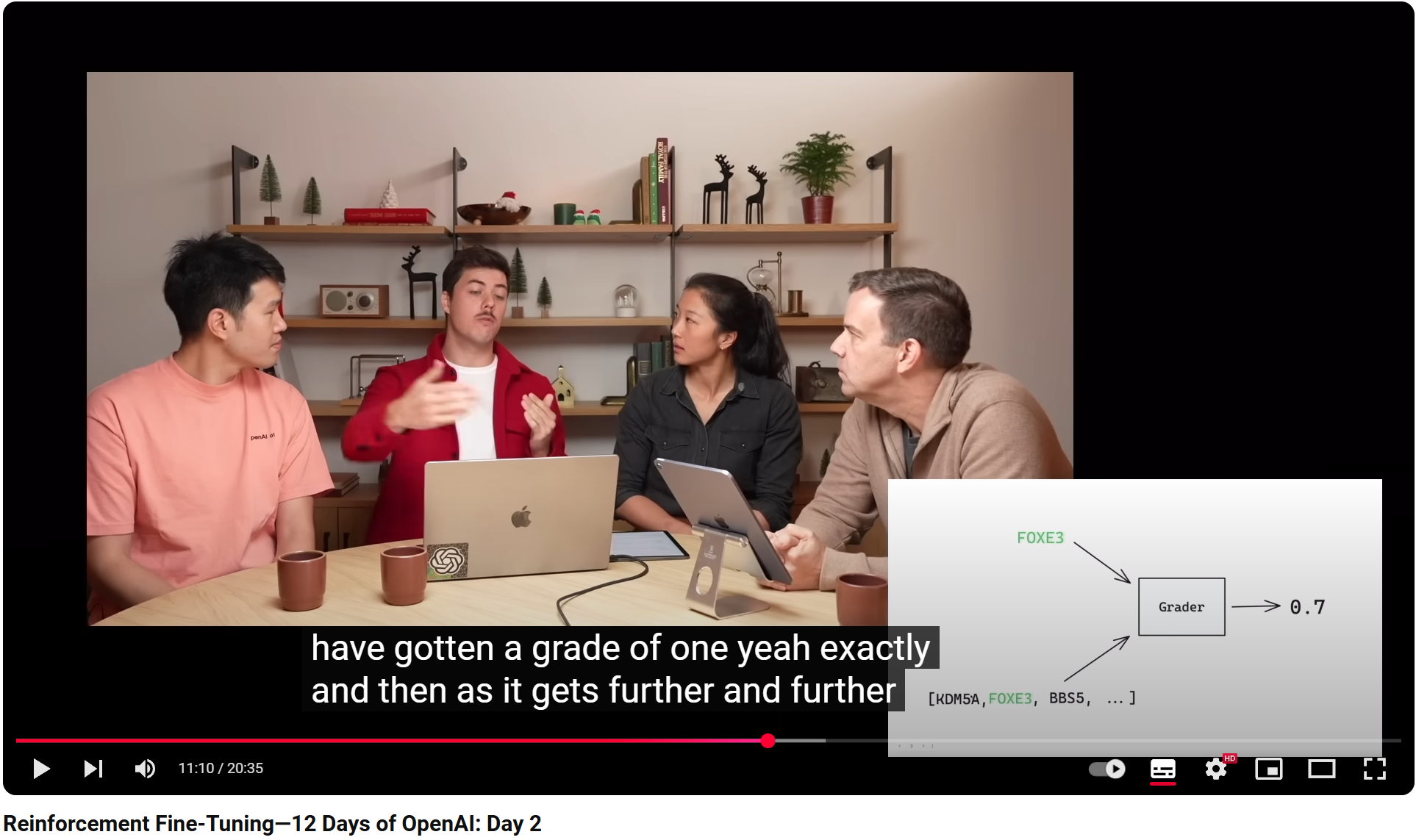}} \\
\end{tabular}

\\

\bottomrule

\end{tabularx}
\vspace{-0.2in}
\end{table*}

To comprehensively evaluate \model's capabilities, we conduct a detailed case study examining its response to a specific query: ``\emph{the role of graders in reinforcement fine-tuning}'', drawn from OpenAI's landmark 12-day video series released in late 2024 [as educational materials]. The query's target information is primarily located in Day 2's content, which details OpenAI's systematic and innovative approach to model enhancement through reinforcement fine-tuning techniques.

\noindent \textbf{Retrieval Accuracy and Response Quality}.
Table~\ref{tab:case study} presents \model's response alongside its retrieved video clips. Our analysis reveals that \model successfully identified and extracted relevant content from Day 2, specifically focusing on reinforcement fine-tuning discussions within the broader context of the 12-video series. The retrieved two-minute clips comprehensively cover: (1) Fundamental concepts of graders; (2) Operational mechanisms of the grading system; (3) Practical examples of partial credit assignment. Within the table, we highlight portions of \model's response that directly correspond to the retrieved video clips. This visualization demonstrates how \model leverages retrieved information to construct detailed and well-supported answers.

\noindent \textbf{Comparative Performance}. A comparative analysis with LightRAG (detailed in Appendix~\ref{apd:case study}) reveals performance distinctions in handling technical content. While both models successfully convey the core concepts of ``\emph{the grading system}'' in reinforcement learning, LightRAG's response demonstrates notable limitations in granularity and scope. Specifically, LightRAG's output lacks crucial technical elements in explaining ``\emph{grader scoring mechanisms}''. Although LightRAG's response maintains fundamental accuracy, it falls short of the comprehensive depth and technical precision exhibited by \model, which provides a nuanced explanation of the grading system's intricacies.

\noindent \textbf{Key Findings}. This case study provides compelling evidence of \model's effectiveness in three critical areas: its ability to construct precise knowledge graph structures that capture complex relationships, its successful leverage of multi-modal information for highly accurate content retrieval, and its enhanced capability to process and synthesize information from multiple long-context videos. These capabilities collectively demonstrate \model's advanced proficiency in handling sophisticated multi-modal tasks while maintaining high standards of accuracy and relevance.

\section{Related Work}

\textbf{Retrieval-Augmented Generation}. 
RAG has emerged as a pivotal paradigm in elevating performance of LLMs. By seamlessly integrating relevant information retrieved from external databases, these systems are able to ground their responses in rich, factual, and domain-specific knowledge~\cite{LightRAG, MemoRAG, RAGSurvey}. At the core of the RAG process lie three essential components: indexing, retrieval, and generation. First, raw data is meticulously processed and structured into a comprehensive database. Next, this database is intelligently queried to retrieve most pertinent information based on user inputs. Finally, this retrieved knowledge is leveraged to generate informed and insightful responses.

Recent advancements in RAG have followed two distinct methodological trajectories. Chunk-based approaches~\cite{NaiveRAG, ChunkRAG, RQ-RAG} have focused on optimizing text segmentation and retrieval through advanced vector space embeddings. In parallel, graph-based methods~\cite{GraphRAG, LightRAG, SubgraphRAG} have explored the use of structured knowledge representations to enhance the efficiency and precision of the retrieval process. Concurrent to these text-centric innovations, the research community has also made significant strides in developing multi-modal RAG systems~\cite{VisRAG, ColPali, MM-VID}, leveraging databases as rich, multi-faceted documents to enrich the knowledge retrieval and generation capabilities.

However, one crucial medium of knowledge remains relatively underexplored in the context of RAG – videos. Preliminary efforts, such as MM-VID~\cite{MM-VID} and iRAG~\cite{iRAG}, have taken initial steps to bridge this gap, but substantial challenges remain in effectively organizing and extracting video-based knowledge. This is where the VideoRAG framework stands as a groundbreaking innovation. By synthesizing state-of-the-art techniques from both text-based and multi-modal RAG approaches, \model\ constructs a comprehensive knowledge graph that seamlessly integrates knowledge from multiple long-form video sources. Coupled with its multi-modal retrieval matching capabilities, \model\ empowers LLMs to tap into the wealth of information inherent in video content, elevating their ability to provide accurate, informed, and contextually relevant responses to user queries.

\textbf{Long Video Understanding}
Extracting meaningful knowledge from long-context videos poses a challenge in the domain of video understanding. Traditional approaches, such as large video language models (LVLMs), have made significant strides by converting video frames into vision tokens for comprehension by large language models~\cite{LongVideoBench, LVBench, HourVideo, Llama-VID, Video-XL, LongVA, INTP}. However, as video length and quantity increase, the computational demands also escalate, necessitating a balance between video length and available resources. This has motivated the exploration of more efficient and scalable solutions to address the growing need for effective long-video understanding.

To this end, our innovative approach presents a novel framework that leverages an efficient video language model for video segment-based general knowledge extraction and query-specific information retrieval. By constructing a graph that integrates information from multiple videos with visual features, we enhance the precision and comprehensiveness of query responses, while accommodating input videos of arbitrary lengths and quantities. This contrasts with existing agent- or retrieval-augmented generation-based methods~\cite{VideoAgent, Video-Agent, Video-RAG, DrVideo}, which rely heavily on external tools for frame-level information extraction, limiting their capacity to respond to diverse queries due to the inherent constraints of these tools. Our holistic approach, which seamlessly combines efficient video understanding with advanced knowledge organization and retrieval, represents a significant advancement in the field, poised to unlock new possibilities in long-video comprehension and query-answering. 
\section{Conclusion}

This paper introduces \model, a novel retrieval-augmented generation framework designed for understanding extremely long-context videos. Through a dual-channel architecture that seamlessly integrates graph-based textual knowledge grounding with multi-modal context encoding, \model effectively processes, indexes, and retrieves information from unlimited-length videos for large language model enhancement. Our comprehensive empirical evaluation on the established LongerVideos benchmark demonstrates \model's superior performance compared to existing RAG alternatives and long video understanding methods across multiple dimensions. The framework's demonstrated capabilities—constructing precise video knowledge structures, leveraging multi-modal information for accurate content retrieval, and processing information from multiple long-context videos—showcase its significant potential for advancing video-based knowledge retrieval and generation tasks.

\clearpage

\bibliographystyle{unsrtnat}
\bibliography{neurips_2024}

\appendix
\clearpage
\section{Details of LongerVideos}\label{apd:data}
LongerVideos is a comprehensive benchmark dataset designed to evaluate a model's ability to comprehend and extract knowledge from long-form videos. By leveraging semantic connections across multiple sources, the dataset facilitates the development of efficient, knowledge-based question-answering. The core methodology presents the model with diverse video lists of varying lengths, then assesses the model's output in terms of completeness, accuracy, and diversity. This holistic approach ensures the evaluated models demonstrate a robust understanding of the content, the ability to synthesize information, and the aptitude to generate well-rounded responses.
\vspace{-0.05in}
\begin{itemize}[leftmargin=*]
    \item \textbf{Input Data}: A diverse collection of long videos, with durations ranging from minutes to hours.
    \item \textbf{Question}: A series of open-ended questions carefully tailored to the provided video list.
    \item \textbf{Expected Output}: Individual responses generated based on the information extracted from videos.
\end{itemize}
\vspace{-0.05in}
The LongerVideos dataset was constructed by systematically curating diverse video lists from YouTube, leveraging the platform's structure where creators often compile thematic content. A major data source comprised online course videos, typically segmented into multiple recordings corresponding to distinct course chapters. For each video, the team employed the yt-dlp tool to download the content in 720P resolution, after which they prepared open-ended questions for each list with the assistance of NotebookLM, a robust multi-video understanding model from Google that can process various videos as input to generate relevant answers. The final LongerVideos dataset consists of 22 carefully curated video lists, with detailed statistics provided in Table~\ref{tab:detail stats}.
\begin{table}[h]
\vspace{-0.05in}
\centering
\caption{Detailed statistics of the \textit{LongerVideos} dataset.}
\label{tab:detail stats}
\resizebox{1.0\textwidth}{!}{
\begin{tabular}{@{}l|lcccc@{}}
\toprule
Video Type & video list name & \#video & \#query & \#overall duration \\
\midrule
\multirow{12}{*}{\textbf{Lecture}}
& \texttt{climate-week-at-columbia-engineering} & 4 & 26 & 5.91 hours \\
& \texttt{rag-lecture} & 19 & 38 & 5.34 hours \\
& \texttt{ai-agent-lecture} & 39 & 45 & 9.35 hours \\
& \texttt{daubechies-wavelet-lecture} & 4 & 25 & 8.97 hours \\
& \texttt{daubechies-art-and-mathematics-lecture} & 4 & 21 & 4.87 hours \\
& \texttt{tech-ceo-lecture} & 4 & 31 & 4.83 hours \\
& \texttt{dspy-lecture} & 9 & 38 & 4.22 hours \\
& \texttt{trading-for-beginners} & 2 & 23 & 4.11 hours \\
& \texttt{ahp-superdecision} & 11 & 24 & 2.40 hours \\
& \texttt{decision-making-science} & 4 & 26 & 2.20 hours \\
& \texttt{12-days-of-openai} & 12 & 35 & 3.43 hours \\
& \texttt{autogen} & 23 & 44 & 8.70 hours \\
\midrule
\multirow{5}{*}{\textbf{Documentary}}
& \texttt{fights-in-animal-kingdom} & 1 & 11 & 3.00 hours \\
& \texttt{nature-scenes} & 1 & 17 & 3.98 hours \\
& \texttt{education-united-nations} & 6 & 39 & 8.41 hours \\
& \texttt{elon-musk} & 1 & 13 & 8.63 hours \\
& \texttt{jeff-bezos} & 3 & 34 & 4.47 hours \\
\midrule
\multirow{5}{*}{\textbf{Entertainment}}
& \texttt{black-myth-wukong} & 10 & 23 & 21.36 hours \\
& \texttt{primetime-emmy-awards} & 3 & 17 & 7.31 hours \\
& \texttt{journey-through-china} & 1 & 27 & 3.37 hours \\
& \texttt{fia-awards} & 1 & 27 & 3.02 hours \\
& \texttt{game-awards} & 2 & 18 & 6.73 hours \\
\bottomrule
\end{tabular}
\vspace{-0.05in}
}
\end{table}

\section{Details of Case Study}\label{apd:case study}
This section provides further details on the case study presented in Section 2, which investigates the purpose and functionality of 'graders' in the context of reinforcement fine-tuning. The investigation utilizes input from the "12 Days of OpenAI" video series, comprising 12 videos that showcase OpenAI's activities in late 2024. To effectively answer the question, the model retrieves relevant content that specifically discusses the role of graders within the reinforcement fine-tuning context. To further illustrate our model's capabilities in retrieving detailed information from videos for generating nuanced answers, we also present a response from another retrieval-augmented generation model, LightRAG, for comprehensive analysis. A comparison of the generated answer by our model, as shown in Table 1, reveals that it provides greater detail, including the scoring criteria used by graders and specific examples illustrating the partial scores assigned. Moreover, it delineates the processes through which graders operate in reinforcement learning, thereby enhancing the overall quality of the generated response. Consequently, the win-rate comparison with GPT-4o-mini emphasizes the empowerment and trustworthiness of the response generated by our \model.
\vspace{-0.05in}
\begin{itemize}[leftmargin=*]

    \item \textbf{Empowerment}: Our \model\ empowers the reader by explaining how graders work in a practical sense, including details on scoring, input comparison, and their role in the feedback loop. This allows readers to grasp how to utilize graders effectively in machine learning contexts.
    
    \item \textbf{Trustworthiness}: Our \model\ aligns closely with established concepts in machine learning and provides more operational details, enhancing credibility. It clearly explains the scoring method and implications for model performance, reinforcing its trustworthy nature.
\end{itemize}
\vspace{-0.1in}

\begin{table*}[t]
\centering
\caption{Further details on the case study concerning the responses of LightRAG to the same question.}
\label{tab:case study}
\small
\begin{tabularx}{\textwidth}{X X}

\toprule
\textbf{Query:} Explain the purpose and functionality of 'graders' in the context of reinforcement fine-tuning.\\

\midrule
\textbf{Input Videos:} \href{https://www.youtube.com/playlist?list=PLOXw6I10VTv9lin5AzsHAHCTrC7BdVdEM}{12 Days of OpenAI} (A total of 12 videos) \\

\midrule

\textbf{LightRAG Response:} 

\texttt{\#\#\# Purpose and Functionality of Graders in Reinforcement Fine-Tuning}

\texttt{Graders serve a crucial role within the context of reinforcement fine-tuning, particularly focusing on evaluating and improving the performance of machine learning models. These evaluative tools assist in determining how well a model aligns with desired outcomes based on specific criteria.}

\\

\texttt{\#\#\#\# Evaluation Process}

\texttt{In reinforcement fine-tuning, graders provide a quantitative measure for assessing the effectiveness of a model's outputs. They help in evaluating the responses generated by the AI against predefined correct answers or standards. This comparison is essential as it offers insights into areas where the model performs well and where it may require further improvements.}

\texttt{The process involves analyzing the outputs of the model after it has undergone reinforcement learning techniques. By examining the consistency and accuracy of the generated responses, graders facilitate the identification of discrepancies and help in recalibrating the model to optimize its performance further. } 

\\

\texttt{\#\#\#\# Role in Feedback Loop}

\texttt{Grading is also integral to the feedback mechanism within reinforcement fine-tuning. The insights gained from the grading process inform the adjustments necessary for the model's reinforcement learning parameters. In this way, graders contribute to refining the model’s learning algorithms, effectively guiding the subsequent training iterations.}

\texttt{Overall, the implementation of graders in reinforcement fine-tuning exemplifies an iterative improvement strategy, where ongoing assessment and feedback drive enhancements in model performance. This systematic approach not only improves the quality of outputs but also ensures that the model continues evolving to meet dynamic user and operational requirements.} 

\\

\bottomrule

\end{tabularx}
\vspace{-0.2in}
\end{table*}

\section{Instructions for Win-Rate \& Quantitative Performance Comparison}\label{apd:prompt4evaluation}

\begin{figure*}[t]
\centering
\begin{tcolorbox}[title=\texttt{Instructions for Win-Rate \& Quantitative Comparison}]
\textbf{Win-Rate Comparison\\}

You will evaluate two answers to the same question based on these criteria: **Comprehensiveness**, **Empowerment**, **Trustworthiness**, **Depth** and **Density**.\\

- **Comprehensiveness**: How much detail does the answer provide to cover all aspects and details of the question?\\
- **Empowerment**: How well does the answer help the reader understand and make informed judgments about the topic?\\
- **Trustworthiness**: Does the answer provide sufficient detail and align with common knowledge, enhancing its credibility?\\
- **Depth**: Does the answer provide in-depth analysis or details, rather than just superficial information?\\
- **Density**: Does the answer contain relevant information without less informative or redundant content?\\
For each criterion, choose the better answer (either Answer 1 or Answer 2) and explain why. Then, select an overall winner based on these criteria.\\

Here is the question:
\texttt{\{query\}\\}
\texttt{Here are the two answers:\\}
**Answer 1:**
\texttt{\{answer1\}\\}
**Answer 2:**
\texttt{\{answer2\}\\}

Evaluate both answers using the criteria listed above and provide detailed explanations for each criterion.
Output your evaluation in the following JSON format:

(The remaining content are omitted for brevity.)

\tikz \draw[dashed] (0,0) -- (\linewidth,0);
\textbf{Quantitative Comparison\\}

You are an expert evaluating an answer against a baseline answer based on these criteria: **Comprehensiveness**, **Empowerment**, **Trustworthiness**, **Depth** and **Density**. \\

(We omit the similar part on win-rate comparison here for brevity.)\\

For the evaluated answer labeled "Evaluation Answer," assign a score from 1 to 5 for each criterion compared to the baseline answer labeled "Baseline Answer." Then, assign an overall score based on these criteria.
The evaluation scores are defined as follows:\\
- 1: Strongly worse than the baseline answer\\
- 2: Weakly worse than the baseline answer\\
- 3: Moderate compared to the baseline answer\\
- 4: Weakly better than the baseline answer\\
- 5: Strongly better than the baseline answer\\

Here is the question:
\texttt{\{query\}}

Here are the answers:

**Baseline Answer:**
\texttt{\{baseline\_answer\}}

**Evaluation Answer:**
\texttt{\{evaluation\_answer\}\\}

Evaluate the answer using the criteria listed above and provide detailed explanations for the scores.
Output your evaluation in the following JSON format:

(The remaining content are omitted for brevity.)

\end{tcolorbox}
\caption{Instructions for LLM-based answer comparison and scoring}
\label{fig:prompt}
\end{figure*}

We present the instructions employed for LLM-based evaluation in Figure~\ref{fig:prompt}, which includes both win-rate comparison and quantitative comparison. For the win-rate comparison, we input the query alongside two competing answers, designated as \texttt{answer1} and \texttt{answer2}, while alternating their positions across multiple iterations to mitigate any positional bias affecting LLM inference.

In the quantitative comparison, we leverage a standard answer from NaiveRAG~\cite{NaiveRAG} labeled \texttt{baseline\_answer}, against which the evaluated answer, referred to as \texttt{evaluation\_answer}, is assessed. The LLM assigns a score from 1 to 5, indicating whether the evaluated answer is inferior or superior to the baseline. This instruction allows us to compare the outputs of multiple models against the same standard answer, thus eliminating the need to adjust their positions. Since all methods are evaluated consistently against the same baseline, positional bias is inherently mitigated, enabling a direct comparison of scores across different methods.

\end{document}